\documentclass[12pt]{iopart}

\usepackage{xcolor} 
\usepackage{multirow}
\usepackage{iopams}
\usepackage{graphicx}
\usepackage{booktabs}

\usepackage[numbers,square,sort&compress]{natbib}

\usepackage[colorlinks=true,linkcolor=blue,citecolor=blue,urlcolor=blue]{hyperref}
\begin{document}

\title[]{Benchmarking machine-learned interatomic potentials for molecular infrared spectroscopy}

\author{Nitik Bhatia$^{1,2,3,4}$, Ond\v{r}ej Krej\v{c}\'{i}$^{2,5,6}$, and Patrick Rinke$^{1,2,3,4,*}$}
\address{$^1$ Department of Physics, Technical University of Munich, James-Franck-Strasse 1, Garching, 85748, Germany}
\address{$^2$ Department of Applied Physics, Aalto University, P.O. Box 11000, AALTO, FI-00076, Finland}
\address{$^3$ Munich Center for Machine Learning (MCML), Munich, Germany}
\address{$^4$ Atomistic Modelling Center, Munich Data Science Institute, Technical University of Munich, Walther-Von-Dyck Str. 10, Garching, 85748, Germany}
\address{$^5$ Department of Mechanical and Materials Engineering, Vesilinnantie 5, Turku, Finland}
\address{$^6$ Department of Chemistry and Material Science, Aalto University, P.O. Box 11000, AALTO, FI-00076, Finland}

\address{$^*$ Author to whom any correspondence should be addressed.}
\ead{patrick.rinke@tum.de}

\vspace{10pt}
\begin{indented}
\item[]Supplementary material for this article is available online.
\end{indented}

\begin{abstract}
Machine learning has transformed the field of atomistic simulations by enabling the development of interatomic potentials that are computationally efficient and highly accurate. These advances have opened the door to modeling molecular vibrations and predicting infrared spectra with near–\textit{ab-initio} accuracy at a fraction of the computational cost. Among these approaches, message-passing neural networks (MPNNs) have emerged as a particularly powerful class of models for representing complex atomic interactions.
In this study, we benchmark five MPNN architectures, SchNet, FieldSchNet, SO3Net, PaiNN, and MACE, for predicting infrared spectra of small organic molecules. SchNet and FieldSchNet are invariant models, while SO3Net, PaiNN, and MACE are equivariant, explicitly accounting for rotational symmetries in molecular representations. We evaluate their performance in terms of computational efficiency, accuracy, and robustness. All models accurately predict properties, such as energies, forces, and dipole moments, required for infrared spectra calculations. They also capture harmonic frequencies and infrared spectra derived from molecular dynamics with high fidelity for molecules in the training set. However, SchNet and FieldSchNet show limited transferability to unseen systems, while SO3Net, PaiNN, and MACE generalize more effectively. In terms of computational efficiency, SchNet is the most efficient and FieldSchNet enables field-dependent response modeling but with higher cost. PaiNN achieves the best balance between accuracy and efficiency, MACE provides the highest spectral accuracy and transferability, and SO3Net performs between PaiNN and MACE. 
\end{abstract}
\vspace{0.5pc}

%
%
%
\maketitle
%
%

\section{Introduction}
Infrared (IR) spectroscopy is a powerful analytical technique widely employed to probe molecular structures, reaction mechanisms, and surface interactions across various scientific disciplines \cite{Khan2018, C3CS60374A}. For instance, in catalysis, it serves as a key tool for identifying reaction intermediates and active sites, thereby guiding the rational design of improved catalysts \cite{C3CS60374A, Concepción18}. However, interpreting experimental IR spectra often remains challenging due to complex spectra 
arising from multiple vibrational modes and environmental effects \cite{Interpret_exp_IR, Intepret_exp_IR_2}. 
Theoretical calculations are often required to aid interpretation, with the harmonic approximation being a standard approach \cite{AIMD_paper, xu2024harmonic}. However, the harmonic approximation inevitably omits anharmonic effects, which are critical for accurate spectral predictions \cite{AIMD_paper, Gaigeot2015, bloino2015anharmonic, conte2023anharmonicity}. Theoretical IR spectroscopy using density functional theory (DFT) based \textit{ab-initio} molecular dynamics (AIMD) simulations provide a solution  by capturing changes in the molecular dipole moment essential for IR spectra prediction \cite{Gaigeot2003, AIMD_paper, Gaigeot2015}. Although AIMD does not fully account for nuclear quantum effects such as zero-point energy and tunneling, recent studies \cite{FieldSchNet, Bhatia2025_PALIRS} have demonstrated that classical AIMD approaches can nonetheless reproduce the principal spectral shapes and characteristic features. However, the substantial computational cost of AIMD still limits its applicability to large systems and long simulation times \cite{gastegger_machine_2017}, highlighting the need for more efficient methodologies to achieve accurate and high-quality IR spectra.

Machine-learned interatomic potentials (MLIPs) have emerged as powerful alternatives for modeling atomistic systems and performing molecular dynamics (MD) simulations with near–first-principles accuracy at a fraction of the computational cost \cite{General_MLIP_1, Deringer_2019, Four_generation, NN_potential, ML_H2}. By accurately predicting potential energies and atomic forces, and thereby reconstructing the underlying potential energy surface, MLIPs can effectively reproduce AIMD trajectories, enabling the exploration of molecular structure, dynamics, and thermodynamics over extended time and length scales. This capability positions MLIPs as practical substitutes for AIMD in large-scale or long-timescale simulations, paving the way for efficient modeling of atomistic systems \cite{gastegger_machine_2017, SchNet_2, SchNet_3, PaiNN, FieldSchNet, MACE, Bhatia2025_PALIRS, bhatia2026mace4irmoluncertaintyawarefoundationmodel}.

The development of MLIPs began with the Behler–Parrinello neural networks (BPNNs) \cite{Behler_2007}, which represented the total potential energy as a sum of atomic contributions and successfully extended molecular dynamics beyond the limits of \textit{ab initio} simulations. The BPNN framework was later 
advanced by the Gaussian Approximation Potential (GAP) \cite{bartok_gaussian_2010, GPR}, which introduced the use of atomic forces and uncertainty quantification to support active learning and improve sampling efficiency. 
Both GAP and BPNN relied on predefined local descriptors, mathematical representations of the atomic environment that encode the relative positions of neighboring atoms in a rotation-, translation-, and permutation-invariant way. In BPNNs, these descriptors are atom-centered symmetry functions, while GAP employs the smooth overlap of atomic positions \cite{SOAP, himanen2020dscribe}.
However, these handcrafted descriptors limited scalability and transferability, leading to difficulties in modeling chemically diverse systems or complex reactive environments.

Recent advances in graph neural networks (GNNs) have addressed these challenges by representing atoms as nodes and their interactions as edges, allowing models to learn how atomic environments determine molecular structure and properties. Message-passing neural networks (MPNNs) extend this framework by propagating information over multiple interaction layers, capturing effects beyond nearest neighbors. Importantly, in MPNNs, the descriptor becomes an integral, learned component of the model itself, rather than a predefined input, which significantly improves transferability and generalization across diverse chemical systems. MPNNs  can predict energies, forces, and dipole moments with high fidelity, making them particularly suitable for machine learning (ML)-based IR spectra prediction \cite{gastegger_machine_2017, SchNet_2, SchNet_3, PaiNN, MACE}. A central distinction within MPNNs concerns how geometric transformations are handled. Invariant architectures produce outputs that remain unchanged under rotations or translations, whereas equivariant architectures preserve directional relationships, typically leading to improved predictions of vector-valued quantities such as forces \cite{PaiNN, Batatia2025Design}.

Despite these advances, the diversity of MPNN architectures has introduced new challenges in understanding how architectural choices influence spectral accuracy, model robustness, and computational efficiency. Reported performance differences across models are often difficult to interpret because evaluations are frequently conducted on different datasets, training protocols, or model implementations, limiting the comparability of published results. Although standardized benchmark platforms \cite{riebesell2025framework, choudhary2024jarvis, molbench2026} exist for certain materials-property prediction tasks, comparable benchmarking frameworks for molecular IR spectroscopy remain limited. Consequently, systematic evaluations that compare multiple architectures under identical conditions are still rare, making it difficult to isolate how specific architectural choices influence predictive performance. In particular, the relative strengths and limitations of invariant and equivariant formulations in reproducing potential energies, forces, molecular dipole fluctuations, and ultimately IR spectra remain insufficiently understood. Furthermore, it is still unclear how increased architectural complexity translates into improvements in accuracy, robustness, and efficiency for IR spectroscopy applications.

Building on these considerations, we aim to utilize the curated dataset developed in our recent work \cite{Bhatia2025_PALIRS} to systematically evaluate widely used MPNN architectures, including SchNet \cite{SchNet_2, SchNet_3}, FieldSchNet \cite{FieldSchNet}, SO3Net \cite{schutt_schnetpack_2019, SchNetpack_2023}, PaiNN \cite{PaiNN}, and MACE \cite{Batatia2025Design, MACE}. Invariant SchNet  and FieldSchNet rely on scalar representations, while equivariant models such as SO3Net, PaiNN, and MACE incorporate vectorial and tensorial features to account for directional dependencies in atomic interactions. All of these models are capable of predicting energies, forces and dipole moments for IR predictions.
FieldSchNet is distinguished by its field-based architecture, which allows the direct prediction of response properties such as dipole moments, whereas other MPNNs typically learn these quantities as separate outputs.

These architectural differences motivate a systematic assessment of each model’s ability to predict infrared spectra of small organic molecules while balancing accuracy, computational efficiency, and robustness. Our analysis emphasizes the importance of selecting models that align with specific simulation requirements, such as achieving accurate potential energy surfaces and spectral predictions. In addition, we discuss the implications of these findings for advancing the predictive capabilities of MLIPs and their integration into workflows for studying molecular processes across diverse chemical environments. By highlighting the strengths and trade-offs of each model, we aim to provide a framework for the development and application of MLIPs in spectroscopic and molecular simulation studies.

\section{Methods}

\subsection{Computational details}
\subsubsection{Database.}
The benchmark dataset consists of 24 small organic molecules (refer to Figure S1 in Supplementary Information) and includes a total of 16,485 structures. This dataset was generated in our previous work \cite{Bhatia2025_PALIRS} using an active learning scheme. The process involved performing ML-driven MD simulations, picking structures that are not well represented by the models trained on the current database, adding these new structures to the dataset, and subsequently retraining the models. The structures were sampled at three different temperatures: 300 K, 500 K, and 700 K. DFT calculations were performed at the PBE \cite{perdew_generalized_1997} level, combined with the Tkatchenko–Scheffler treatment of van der Waals interactions \cite{tkatchenko_accurate_2009} and ``light'' basis sets as implemented in the \textsc{FHI-aims} \cite{blum2009ab, havu2009efficient, levchenko2015hybrid,Xinguo/implem_full_author_list} package. Further details regarding the dataset and its generation can be found in our previous work \cite{Bhatia2025_PALIRS}.

\subsubsection{Test sets for model assessment.}
For model validation, multiple test sets were employed. First, the benchmark data set (referred to as the AL test set) described above was divided into training, validation, and test sets. Additionally, we make use of an independent test set generated in our earlier work \cite{Bhatia2025_PALIRS}, hereafter referred to as the PALIRS-MD (PMD) test set. This PMD test set was constructed from 100 ps molecular dynamics simulations at 300 K and comprises 480 structures in total, with 20 representative structures per molecule selected using constrained K-means clustering \cite{bradley2000constrained}. 

Finally, to benchmark inference performance, two additional methanol test sets were derived from AIMD trajectories at 300 K and 700 K, each containing 1000 structures. The average inference time per energy and force evaluation was estimated from the total time required to process all 1000 structures at each temperature.

\subsection{Message-passing neural networks}
Message-passing neural networks form the basis of the MLIPs benchmarked in this work. In these MLIPs, atomic configurations are represented as graphs, where atoms correspond to nodes and edges connect neighboring atoms within a predefined cutoff distance. Information is exchanged between atoms through iterative message-passing and update steps, repeated up to a maximum number of interaction layers 
$\mathit{t}_{max}$. This procedure yields atomic representations that capture both local environments and higher-order many-body interactions beyond the cutoff distance \cite{ML_H2, Stark_2024}. The general message-passing formalism can be expressed as:

\begin{equation}
    \mathbf{m}^{\,t+1}_{i} = \sum_{j \in \mathcal{N}(i)} \mathbf{M}_{t} \boldsymbol{\big(} \mathbf{s}^{t}_{i}, \mathbf{s}^{t}_{j}, \mathbf{e}_{ij} \boldsymbol{\big)}, \label{eq11}
\end{equation}

\begin{equation}
    \mathbf{s}^{t+1}_{i} = \mathbf{U}_{t} \boldsymbol{\big(} \mathbf{s}^{t}_{i}, \mathbf{\vec{m}}^{\,t+1}_{i} \boldsymbol{\big)}, \label{eq2} 
\end{equation} 
where $\mathbf{m}^{\,t+1}_{i}$ is the message aggregated from neighboring nodes $j \in \mathcal{N}(i)$ at step $t$, using the scalar atomic features $\mathbf{s}^{t}_{i}$ and $\mathbf{s}^{t}_{j}$, which are connected by the edge features $\mathbf{e}_{ij}$. 
The edge features $\mathbf{e}_{ij}$ are typically defined as a function of the interatomic distance $r_{ij}$.
$\mathbf{M}_{t}$ and $\mathbf{U}_{t}$ can be linear or nonlinear message and update functions, respectively \cite{MPNN-gilmer17a}. 
Both SchNet and its extension, FieldSchNet, are examples of non-equivariant MPNNs. While SchNet focuses on scalar atomic features, FieldSchNet extends this framework by incorporating atomic dipole features, enabling it to model response properties and solvent effects under external fields effectively \cite{schutt_schnetpack_2019, SchNetpack_2023}.

Recent advancements have demonstrated that incorporating equivariant (vectorial) features into MPNNs significantly enhances both the data efficiency and accuracy of these models \cite{PaiNN, Batatia2022Design, NequiP, Allegro}. To integrate these features into the message-passing framework, the message in equation (\ref{eq11}) is modified to:
\begin{equation}
    \mathbf{\vec{m}}^{\,t+1}_{i} = \sum_{j \in \mathcal{N}(i)} \mathbf{\vec{M}}^{\,t} \boldsymbol{\big(} \mathbf{s}^{t}_{i}, \mathbf{s}^{t}_{j}, \mathbf{\hat{v}}^{t}_{i}, \mathbf{\hat{v}}^{t}_{j}, \mathbf{e}_{ij} \boldsymbol{\big)}, \label{eq3}
\end{equation}
where vectorial representations $\mathbf{\hat{v}}$ are additionally included. 
These modifications are implemented in equivariant MPNNs such as SO3Net \cite{SchNetpack_2023}, PaiNN \cite{PaiNN}, and MACE \cite{Batatia2025Design, MACE}, which incorporate vectorial features to represent directional atomic information and capture three-dimensional geometric and many-body interactions.

\subsubsection{MLIP parameters.}
The architectural and training parameters of all benchmarked MLIPs are summarized in Tables~\ref{tab:mlip_standard} and~\ref{tab:mace_parameters}. The SchNetPack-based models, i.e. SchNet, FieldSchNet, SO3Net, and PaiNN, were trained under a unified protocol to ensure a controlled comparison across architectures, sharing identical cutoff distances, training schedules, and loss weighting schemes while differing only in their internal interaction structure.

\begin{table}[ht]
\centering
\caption{Summary of parameters for SchNet, FieldSchNet, SO3Net, and PaiNN models.}
\label{tab:mlip_standard}
\small
\begin{tabular}{lcccc}
\hline
Parameter & SchNet & FieldSchNet & SO3Net & PaiNN \\
\hline
Interaction layers & 6 & 5 & 10 & 10 \\
Message-passing features & 128 & 128 & 128 & 128 \\
Epochs & 1000 & 1000 & 1000 & 1000 \\
Loss weights (E,F,D) & 0.05/0.94/0.01 & 0.05/0.94/0.01 & 0.05/0.94/0.01 & 0.05/0.94/0.01 \\
Cutoff distance (\AA) & 5 & 5 & 5 & 5 \\
Data split & 80/10/10 & 80/10/10 & 80/10/10 & 80/10/10 \\
\hline
\end{tabular}

\vspace{1mm}
\footnotesize Training performed using SchNetPack v2.1.1 (\url{https://github.com/atomistic-machine-learning/schnetpack}).
\end{table}

\begin{table}[ht]
\centering
\caption{Summary of parameters for MACE models: Energy/Force (MACE-EF) and Dipole (MACE-D).}
\label{tab:mace_parameters}
\small
\begin{tabular}{lcc}
\hline
Parameter & MACE-EF & MACE-D \\
\hline
Interaction layers & 2 & 2 \\
Message-passing channels & 256 & 16 \\
Epochs & 1000 & 500 \\
Loss weights (E,F,D) & 1$\rightarrow$10, 10$\rightarrow$1 & -- \\
Cutoff distance (\AA) & 5 & 5 \\
Data split & 85/5/10 & 85/5/10 \\
\hline
\end{tabular}

\vspace{1mm}
\footnotesize Training performed using MACE v0.3.8 (\url{https://github.com/ACEsuit/mace}).
\end{table}

For the MACE framework, two specialized models were trained: an energy–force model (MACE-EF) and a dipole model (MACE-D). Both employ correlation order 3 with two equivariant interaction layers and the same spatial cutoff. The primary distinction lies in their channel capacity and architectural scale: MACE-EF uses a larger hidden representation to prioritize force accuracy, whereas MACE-D adopts a reduced representation tailored to efficient dipole prediction. The corresponding hidden irreducible representations are $256\times0e + 256\times1o + 256\times2e$ for MACE-EF and $16\times0e + 16\times1o + 16\times2e$ for MACE-D, reflecting this trade-off between accuracy and cost. During training of MACE-EF, the relative weights of the energy and force terms in the loss function were adjusted: initially set to 1:10 to emphasize accurate force learning, and after 800 epochs switched to 10:1 to fine-tune energy predictions. This staged weighting strategy helps balance the learning of forces and energies for optimal model performance.

\subsubsection{Harmonic frequency calculations.} 
To benchmark the MLIPs, harmonic vibrational frequencies were computed for all 24 molecules considered in this work. 
Harmonic frequencies are obtained by constructing and diagonalizing the mass-weighted Hessian matrix, whose elements are the second derivatives of the potential energy with respect to atomic displacements. The eigenvalues of this matrix correspond to the squared vibrational frequencies, and the associated eigenvectors define the normal modes of vibration \cite{harmonic_1, Harmoni_2, AIMD_paper}.
For the finite-difference calculation of the Hessian, the atomic displacement was set to 0.002~\AA. All calculations were performed using the Atomic Simulation Environment (ASE 3.22.1) \cite{ase-paper}.

\subsubsection{Molecular dynamics-based IR spectra.}
In this work, the IR spectra are computed using a MD-based approach. First, MD simulations are performed using the MLIP to obtain the energies and forces along the trajectory, from which the dipole moment of each configuration is predicted. 
The IR spectrum is then obtained from the autocorrelation function of the time derivative of the dipole moment ($\dot{\mu}$), expressed as:

\begin{equation}
    I_{IR} \propto \int_{-\infty }^{+\infty}\left\langle \dot{\mu}\left( \tau \right) \dot{\mu}\left( \tau + t \right) \right\rangle_{\tau} e^{-i\omega t} dt. \label{eq:IR-intensity}
\end{equation}

The autocorrelation functions are computed using the Wiener–Khinchin theorem \cite{wiener_generalized_1930}. To ensure high-quality IR spectra, a Hann window function \cite{blackman_measurement_1958} and zero-padding are applied prior to the Fourier transform. A maximum correlation depth of 1000~fs is used. All IR spectra processing was performed using adapted code from SchNetPack \cite{SchNetpack_2023}.

For comparison, reference DFT-based AIMD trajectories and the corresponding IR spectra were taken from our previous work \cite{Bhatia2025_PALIRS}, where the computational setup and parameters are described in detail.
All MLIP-based MD simulations were carried out using the Langevin thermostat \cite{ceriotti_langevin_2009} as implemented in ASE, with a friction coefficient of 0.01. Each simulation ran for a total of 55~ps, using a time step of 0.5~fs. The initial 5~ps were discarded to ensure proper thermalization. Unless otherwise specified, all simulations were conducted at 300~K.

\subsubsection{Similarity measures.}
The agreement between theoretical IR spectra (from MLIPs or AIMD) and experimental spectra is quantified using two metrics: the Pearson's correlation coefficient (PCC) and the Wasserstein distance (WD). The PCC is defined as:

\begin{equation}
    \mathrm{PCC} = \frac{\sum_{i=1}^{n} (x_i - \bar{x})(y_i - \bar{y})}{\sqrt{\sum_{i=1}^{n} (x_i - \bar{x})^2} \sqrt{\sum_{i=1}^{n} (y_i - \bar{y})^2}}, \label{eq:PCC}
\end{equation}
where \(x_i\) and \(y_i\) represent the intensities of the respective spectra, and \(\bar{x}\) and \(\bar{y}\) are their mean values \cite{esch_quantitative_2021, henschel_theoretical_2020, pracht_comprehensive_2020}.

The WD is given by:
\begin{equation}
WD(\mu, \nu) = \inf_{\gamma \in \Gamma(\mu, \nu)} \int_{\mathbb{R} \times \mathbb{R}} ||x - y|| \, d\gamma(x,y), \label{eq:WD}
\end{equation}
where \(\mu\) and \(\nu\) are the distributions corresponding to the two spectra being compared, \(\Gamma(\mu, \nu)\) denotes the set of all joint distributions with marginals \(\mu\) and \(\nu\), \(||x - y||\) is the distance between points sampled from the distributions, and \(d\gamma(x,y)\) represents the infinitesimal mass transported from \(x\) to \(y\) \cite{esch_quantitative_2021, rubner_earth_2000}.
The PCC ranges from -1 to 1, where a value of 1 indicates perfect similarity between the spectra.  
The WD is non-negative, with lower values indicating closer agreement between spectra and a value of 0 corresponding to perfect similarity.

For experimental spectra, preprocessing is required to remove baseline artifacts and match the resolution to theoretical spectra. We applied the automated baseline correction procedure described in \cite{esch_quantitative_2021}, with further details provided in the Supplementary Information of \cite{Bhatia2025_PALIRS}. Linear interpolation was used to align the frequency points of the theoretical spectra to those of the experimental spectra.

\section{Results}
\subsection{Accuracy and efficiency of ML models}
The performance of ML models in terms of both accuracy and efficiency is critical for their applicability in atomistic simulations and IR spectra prediction. In this section, we assess the predictive accuracy of the models, alongside their computational efficiency in terms of training time and resource consumption.

In Table~\ref{accuracy_ML}, we present the mean absolute error (MAE) for energy, force, and dipole moment predictions obtained from the trained models on the AL test set. For energy predictions, all five models exhibit strong performance, with MAE values below 3.5~meV. Among them, the equivariant MPNNs achieve the lowest errors overall, with MACE and PaiNN providing the most accurate energy predictions. Similarly, for force predictions, equivariant MPNNs clearly outperform their non-equivariant counterparts. While the difference in energy MAE is below 1~meV, the force MAE shows a larger disparity: non-equivariant models exceed equivariant ones by approximately 6~meV/Å, roughly doubling the error observed for MACE and PaiNN. 
For dipole moment predictions, the same trend is observed. Equivariant MPNNs again perform best, with PaiNN yielding the lowest error and MACE following closely behind.

\begin{table}[htbp]
\caption{\label{accuracy_ML}Mean absolute error (MAE) in energy, force, and dipole moment of the trained models on the benchmark test dataset, together with time taken for training each model. Bold represents the best-performing models and italics represents the second best.}
\footnotesize
\begin{tabular*}{\textwidth}{@{\extracolsep{\fill}} l l c c c c cc @{}}
\br
Target & Unit & SchNet & FieldSchNet & SO3Net & PaiNN & \multicolumn{2}{c}{MACE} \\
\cmidrule(lr){7-8}
       &      &        &             &        &       & MLIP & Dipole \\
\mr
Energy & meV      & 3.3 & 3.1 & 2.4 & \textbf{2.2} & \textit{2.3} & - \\
Force  & meV/Å    & 13.6 & 12.6 & \textbf{5.8} & \textit{5.9} & \textbf{5.8} & - \\
Dipole moment & mDebye & 13.4 & 22.1 & 13.9 & \textbf{11.0} & - & \textit{13.2} \\
\mr
Time/GPUs & hours & \textbf{18.5/1} & 28.6/1 & 36.1/1 & \textit{28.1/1} & 4.8/8 & 7.5/1 \\
\br
\end{tabular*}
\end{table}

We further evaluated the trained models on the PMD test set as well as on methanol AIMD data at 300 K and 700 K. The MAEs in energy, force, and dipole moment are reported in Tables S1–S3, respectively. In all cases, the models showed excellent accuracy in predicting energies. For forces, the errors obtained with SchNet and FieldSchNet were approximately twice as large as those predicted by SO3Net, PaiNN, and MACE. For dipole moments, all models exhibited comparable performance. Notably, the prediction errors for all models were higher at 700 K compared to 300 K.

To complement this analysis, we further examined the evolution of the absolute error in total energy ($\Delta E = E_{\mathrm{DFT}} - E_{\mathrm{ML}}$) along the 700 K AIMD trajectory of methanol (Figure S2a). Across all models, the error remains consistently small, typically within 1–3 meV, underscoring their robustness. The corresponding heatmap (Figure S2(b)) reveals distinct correlation patterns: SchNet aligns most closely with SO3Net and PaiNN, shows moderate correlation with FieldSchNet, and the weakest with MACE. FieldSchNet correlates strongly with MACE while maintaining moderate agreement with the other models. SO3Net and PaiNN demonstrate the strongest mutual correlation, with both also closely linked to SchNet. By contrast, MACE is most strongly correlated with FieldSchNet, shows intermediate correlation with SO3Net and PaiNN, and the least with SchNet.

To evaluate computational efficiency, we have also reported  the time taken by each of the five ML models,
in the last row of Table \ref{accuracy_ML}. 
As expected, SchNet, being a lightweight and non-equivariant MPNN, demonstrates the highest efficiency, outperforming the other models. On the other hand, FieldSchNet requires a comparable amount of time to the equivariant MPNNs due to the additional computations of atomic dipole. Among the equivariant MPNNs (SO3Net, PaiNN, and MACE), PaiNN emerges as the most computationally efficient model. Both SO3Net and MACE require a similar amount of computational time. 

In addition, we benchmarked the inference time for a single energy and force prediction, estimated from the total time required to evaluate 1000 methanol structures generated from AIMD simulations at 300 K and 700 K (refer to Figure S3 and S4 in the Supplementary Information). In this test, SchNet was the fastest but also the least accurate, MACE was the slowest yet delivered the highest accuracy, while PaiNN offered the best compromise between computational cost and predictive performance. These results highlight the trade-offs between computational efficiency and model complexity across the evaluated models.

\subsection{Performance of ML models in IR spectra prediction}
In this section, we evaluate the performance of models in predicting vibrational-spectroscopic properties.
\subsubsection{Harmonic frequencies prediction.}
We first assess the accuracy of all models in predicting harmonic frequencies for the 24 molecules in the benchmark dataset, providing a test of their ability to capture fundamental vibrational characteristics. The ML predictions are evaluated against DFT reference calculations.

Table~\ref{harmonic_ML} summarizes the MAE of all models. Across the set, SO3Net achieves the lowest MAE of 2.8~cm\(^{-1}\), while PaiNN exhibits the highest MAE of 7.6~cm\(^{-1}\).

\begin{table}[htbp]
\caption{\label{harmonic_ML}Mean absolute error (MAE) in harmonic frequency prediction for the trained models on the benchmark dataset. Bold represents the best-performing models and italics represents the second best.}
\footnotesize
\begin{tabular*}{\textwidth}{@{\extracolsep{\fill}} l l c c c c c @{}}
\br
Target & Unit & SchNet & FieldSchNet & SO3Net & PaiNN & MACE \\
\mr
Harmonic frequency & cm$^{-1}$ & 4.1 & 4.2 & \textbf{2.8} & 7.6 & \textit{3.2} \\
\br
\end{tabular*}
\end{table}

\subsubsection{Molecular dynamics–based IR spectra prediction.}
\begin{figure}[t!]
	\centering 
	\includegraphics[width=1.0\textwidth]{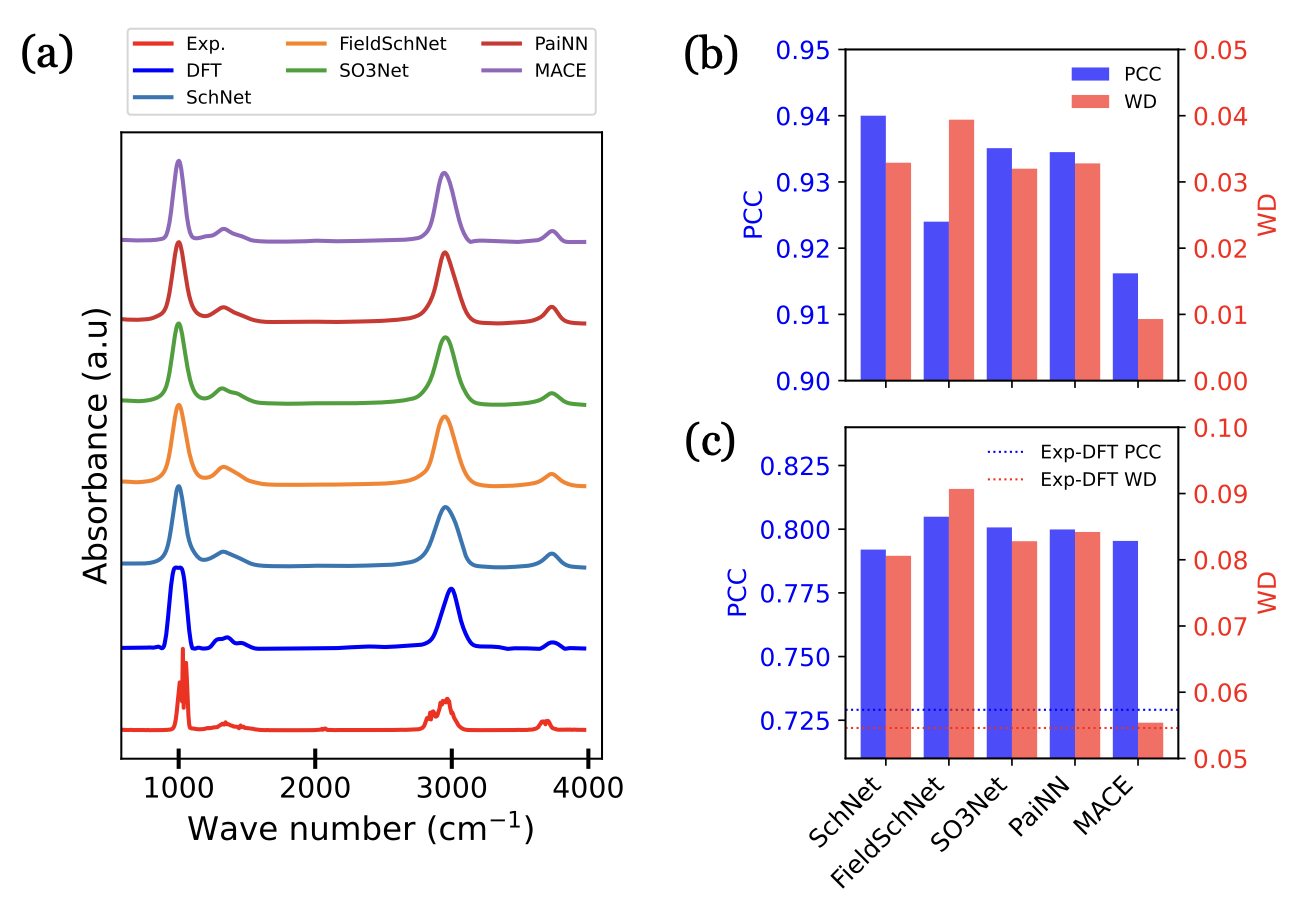}
    \caption{(a) Comparison of IR spectra of methanol at 300 K in the gas phase. (b)-(c) Similarity results for IR spectra prediction of methanol. Pearson Correlation Coefficient (PCC) and Wasserstein Distance (WD) are visualized for: (b) DFT vs. ML models and (c) experiment (Exp) vs ML models with Exp vs. DFT result denoted by dashed lines.}
    \label{Methanol_spectra}
\end{figure}

We next evaluate the models in predicting full IR spectra derived from MD trajectories. As a representative case, methanol is used to benchmark performance.
Figure \ref{Methanol_spectra}(a) illustrates the predicted IR spectra of methanol using the five models alongside the reference spectra from DFT-based AIMD simulations and experimental (Exp) IR spectra extracted from the NIST database \cite{Wallace2024}. Overall, the IR peaks are well reproduced by both the DFT reference and all ML models when compared to the experimental spectra.

To quantify the similarity between the IR spectra, we use PCC and WD, as these measures have been found to be effective in assessing IR spectra similarity \cite{Bhatia2025_PALIRS, esch_quantitative_2021}. Table \ref{similarity_IR} reports the PCC and WD for DFT vs. ML models and Exp vs. ML models. The Exp vs. DFT comparison establishes a baseline for evaluating the ML models. The comparison of ML against DFT predictions reveals exceptionally high PCC values, with SchNet achieving the highest PCC of 0.9400. For WD, MACE significantly outperforms all other models, with a remarkably low WD of 0.0093.

\begin{table}[htbp]
\caption{\label{similarity_IR}Similarity results for IR spectra prediction for methanol in the gas phase. PCC and WD are reported for DFT vs. ML models and Exp vs. ML models. Bold indicates the best-performing models in each category.}
\footnotesize
\begin{tabular*}{\textwidth}{@{\extracolsep{\fill}} l l c c c c c @{}}
\br
Comparison & Metric & SchNet & FieldSchNet & SO3Net & PaiNN & MACE \\
\mr
DFT vs. ML           & PCC & \textbf{0.9400} & 0.9240 & 0.9351 & 0.9345 & 0.9162 \\
                     & WD  & 0.0329 & 0.0394 & 0.0320 & 0.0328 & \textbf{0.0093} \\
\mr
Exp vs. ML  & PCC & 0.7920 & \textbf{0.8049} & 0.8007 & 0.7999 & 0.7954 \\
                     & WD  & 0.0806 & 0.0907 & 0.0828 & 0.0842 & \textbf{0.0554} \\
\br
\end{tabular*}

\vspace{1mm}
\footnotesize{Note that Exp vs. DFT comparison yields PCC = 0.7291 and WD = 0.0546, establishing a baseline for evaluating the ML models.}
\end{table}

Among the ML models in the Exp vs ML comparison, FieldSchNet achieves the highest PCC, followed by SO3Net and PaiNN. While all models surpass the DFT baseline in PCC, MACE emerges as the standout performer with the lowest WD, closely matching the Exp-DFT similarity in terms of spectral alignment.

\begin{figure}[htbp]
	\centering 
	\includegraphics[width=1.0\textwidth]{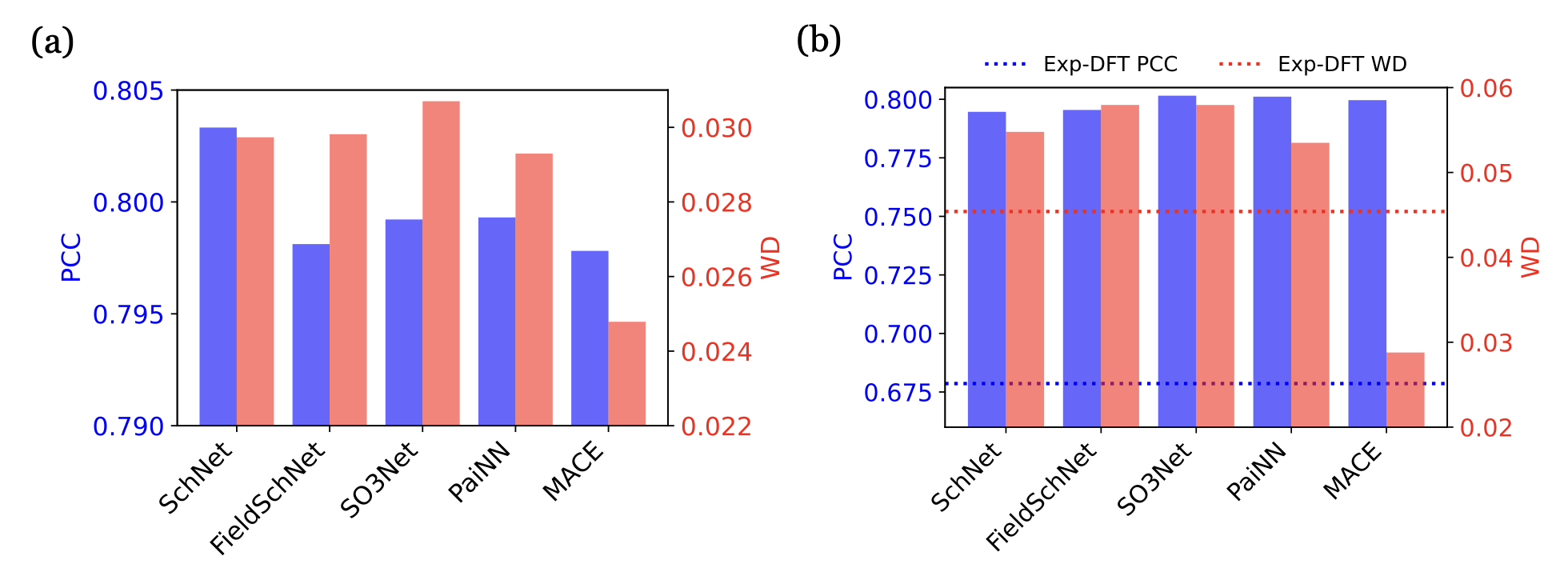}
    \caption{Similarity results for IR spectra prediction for all 24 molecules in gas-phase at 300 K. PCC and WD are reported for (a) DFT vs. ML models, and (b) Exp vs. ML models with Exp vs DFT values denoted by dashed lines.}
    \label{Spectra_metric_overall}
\end{figure}

\begin{table}[htbp]
\caption{\label{performance_24_molecules}Performance statistics for the similarity of IR spectra between Experimental (Exp), DFT, and ML predictions across 24 molecules. The line above a quantity denotes the mean, while $\delta$ represents the standard deviation. Bold indicates the best-performing models for each metric.}
\footnotesize
\begin{tabular*}{\textwidth}{@{\extracolsep{\fill}} l l c c @{}}
\br
Comparison & Model & $\overline{\mathrm{PCC}} \, (\delta \mathrm{PCC})$ & $\overline{\mathrm{WD}} \, (\delta \mathrm{WD})$ \\
\mr
\multirow{5}{*}{DFT-ML} 
                   & SchNet          & \textbf{0.8033 (0.1840)}    & 0.0297 (0.0270) \\
                   & FieldSchNet     & 0.7981 (0.1839)             & 0.0298 (0.0279) \\
                   & SO3Net          & 0.7992 (0.1899)             & 0.0307 (0.0303) \\
                   & PaiNN           & 0.7993 (0.1859)             & 0.0293 (0.0245) \\
                   & MACE            & 0.7978 (0.1870)             & \textbf{0.0248 (0.0225)} \\
\mr
\multirow{5}{*}{Exp-ML} 
                   & SchNet          & 0.7946 (0.1586)             & 0.0548 (0.0202) \\
                   & FieldSchNet     & 0.7954 (0.1578)             & 0.0580 (0.0227) \\
                   & SO3Net          & \textbf{0.8015 (0.1522)}    & 0.0580 (0.0220) \\
                   & PaiNN           & 0.8011 (0.1665)             & 0.0535 (0.0203) \\
                   & MACE            & 0.7996 (0.1618)             & \textbf{0.0288 (0.0140)} \\
\mr
Exp-DFT            & --              & 0.6786 (0.2236)             & 0.0454 (0.0263) \\
\br
\end{tabular*}
\end{table}

Extending the analysis to the 24 molecules in the training set, we evaluated the performance of the ML models in predicting room-temperature (300~K) IR spectra (see Figure \ref{Spectra_metric_overall} and Table \ref{performance_24_molecules}). The comparison between experimental and DFT spectra shows a slightly lower correlation than observed for methanol above, with a mean PCC of 0.6786 and a WD of 0.0454. For both DFT vs ML and experimental vs ML comparisons, all models produced similar PCC values and standard deviations, indicating comparable prediction of peak positions. In contrast, WD values differed more substantially between models. WD values were consistently higher for experimental vs ML than for DFT vs ML comparisons. Among the models, MACE yielded the lowest WD values in both comparisons, while SO3Net produced the highest WD values despite comparable PCC values. 

\subsubsection{ML model performance in IR spectra across temperatures variations.}
To evaluate the robustness of ML models in predicting IR spectra under different temperature conditions, we focus on methanol as a representative case. The performance of the models is assessed by comparing their predictions with DFT-based AIMD spectra at five different temperatures: 100 K, 300 K, 500 K, 700 K, and 900 K. The corresponding spectra are shown in Figure S5 of the Supplementary Information, where all models successfully reproduce key spectral features, including peak intensities and broadening effects, across all five temperatures when compared to the DFT-based AIMD reference.

\begin{table}[htbp]
\caption{\label{temp_variation_ML}Performance of ML models in predicting IR spectra of methanol across five temperatures (100 K, 300 K, 500 K, 700 K, 900 K). The line above a quantity indicates the mean, while $\delta$ represents the standard deviation. The best performance in each metric is highlighted in bold.}
\footnotesize
\centering
\begin{tabular*}{\textwidth}{@{\extracolsep{\fill}} l c c @{}}
\br
Model & $\overline{\mathrm{PCC}} \, (\delta \mathrm{PCC})$ & $\overline{\mathrm{WD}} \, (\delta \mathrm{WD})$ \\
\mr
SchNet        & \textbf{0.9403 (0.0404)} & 0.0038 (0.0016) \\
FieldSchNet   & 0.9386 (0.0359)          & 0.0037 (0.0018) \\
SO3Net        & 0.9381 (0.0344)          & 0.0041 (0.0018) \\
PaiNN         & 0.9396 (0.0357)          & \textbf{0.0037 (0.0018)} \\
MACE          & 0.9124 (0.0656)          & 0.0037 (0.0024) \\
\br
\end{tabular*}
\end{table}

Table \ref{temp_variation_ML} presents the performance of the ML models averaged across five temperatures. As in the case of 300K simulations, all models exhibit nearly identical PCC and WD values, indicating that temperature transfer does not significantly degrade spectral agreement. The small differences between models fall within the observed variability and do not suggest a clear ranking.
MACE shows a slightly lower mean PCC, but its WD remains comparable to the other models, confirming that detailed spectral shapes are preserved. Likewise, the marginal advantage of SchNet in PCC and PaiNN in WD is not statistically meaningful given the overlapping standard deviations. 

A similar analysis was performed for ethanol (Figure S6 and Table S4). In contrast to methanol, the ethanol benchmark shows a clearer separation between models. PaiNN achieves the highest mean PCC and the lowest WD while also exhibiting the smallest variability across temperatures, indicating the most stable transferability. SO3Net and MACE follow closely, with comparable accuracy but slightly larger fluctuations.

\begin{figure}[htbp]
	\centering 
	\includegraphics[width=1.0\textwidth]{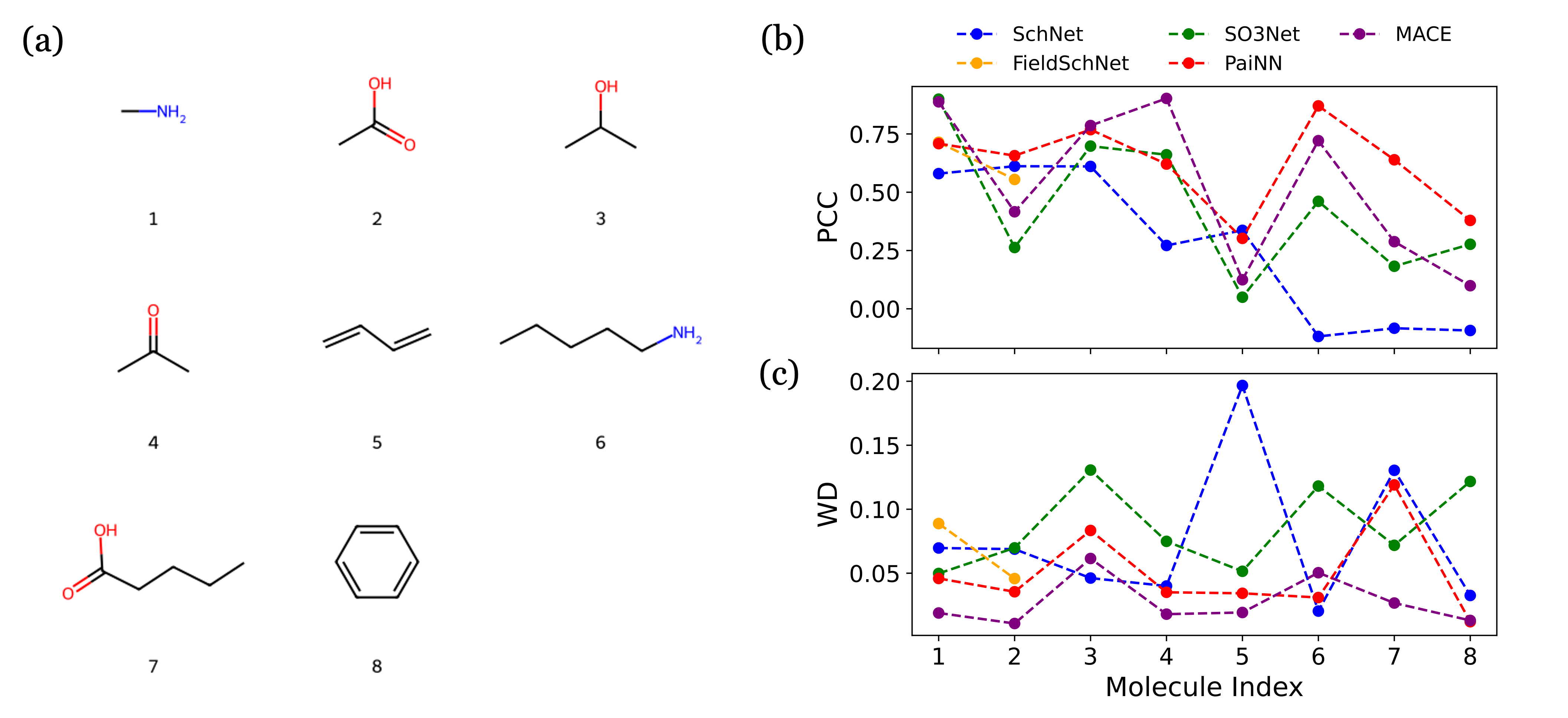}
    \caption{(a) Set of eight representative gas-phase molecules used to evaluate the transferability of the ML models. (b)–(c) Similarity between experimental and ML-predicted (Exp vs. ML) IR spectra quantified by the Pearson correlation coefficient (PCC) and Wasserstein distance (WD), respectively.}
    \label{Transferability_Similarity}
\end{figure}

\subsubsection{Transferability of ML models in predicting IR spectra.}

The transferability of the ML models was evaluated using eight molecules that were not included in the training dataset, spanning a range of molecular sizes and functional motifs (Figure \ref{Transferability_Similarity}(a)). Representative spectra for Molecule 1, Molecule 5, and Molecule 7 are shown in Figure S7 in the Supplementary Information and compared against experimental NIST references.

For the smallest system (Molecule 1), most models achieve PCC values above 0.75, comparable to the agreement observed for molecules within the training set. 
However, a clear degradation in performance is observed as molecular size increases. The invariant architectures show the strongest sensitivity to system complexity, with progressively lower correlation and larger spectral distortions for larger molecules. For instance, FieldSchNet could not be assessed beyond Molecule 2 due to failed MD simulations caused by poor energy and force predictions.
Also, SchNet has the lowest PCC for almost every molecule starting with index 3 and its WD is very often the highest out of all models.
In contrast, the equivariant models retain substantially higher agreement with experiment across the majority of the test systems (Figure \ref{Transferability_Similarity}(b–c)). 

\section{Discussion}

The comprehensive benchmarking of five MPNN architectures reveals a clear trade-off between predictive accuracy, generalization capability, and computational efficiency. Across energy, force, and dipole predictions, equivariant architectures consistently outperform invariant models (refer to Table~\ref{accuracy_ML}), highlighting the importance of incorporating rotationally equivariant representations for learning molecular interactions. The correlation analysis of prediction errors in Figure S2(b) reveals that the models do not behave independently but instead form structured groups with varying degrees of agreement.  
The observed grouping indicates that different architectures do not converge to identical error landscapes, but instead capture partially overlapping yet distinct aspects of the underlying molecular interactions. Extending this perspective to computational considerations, PaiNN emerges as the most balanced architecture overall, achieving strong predictive performance across all properties while maintaining a moderate computational cost (refer to Table~\ref{accuracy_ML}). In contrast, SchNet represents the most computationally efficient option among the tested models, albeit with reduced predictive accuracy, making it more suitable for applications where efficiency is prioritized over maximal fidelity.

Since the ultimate objective of our benchmark study is the reliable prediction of vibrational and spectroscopic observables, model evaluation must extend beyond energies and forces to derived properties directly relevant to IR spectroscopy. We therefore next assess the models in terms of harmonic vibrational frequencies and subsequently through MD-derived infrared spectra. The prediction of harmonic frequencies reveals that accurate energies and forces do not necessarily guarantee accurate derived properties (refer to Table~\ref{harmonic_ML}). Equivariant architectures such as SO3Net and MACE generally offer superior accuracy for second-derivative properties derived from the Hessian, consistent with their lower energy and force errors. However, this relationship is not strict. PaiNN exhibits relatively low force errors yet shows significantly elevated harmonic frequency errors  compared to SO3Net and MACE, demonstrating that second-derivative properties are more sensitive to subtle deficiencies in force predictions than overall error metrics alone suggest. 

The robustness of MD-derived IR spectra contrasts sharply with the sensitivity of Hessian-derived properties. MD-derived spectra show remarkable stability due to averaging effects inherent in trajectory-based calculations, where spectral features integrate over many configurations rather than depending on local curvature at equilibrium geometry.
Across the 24 training molecules at room temperature (300 K), all models show strong agreement with both DFT and experiment, with consistently high PCC values, indicating reliable reproduction of peak positions. In contrast, WD shows clearer model-dependent variation: MACE achieves the lowest value (0.02), while others lie in the range 0.03–0.05, with SO3Net exhibiting comparatively higher WD despite similar PCC, reflecting weaker intensity reproduction. However, the values are still reasonable and consistent with prior studies in the field \cite{esch_quantitative_2021, Bhatia2025_PALIRS}.

While room-temperature spectra provide insight into in-distribution performance, practical spectroscopic applications require models to remain reliable under varying thermodynamic conditions. We therefore evaluate temperature robustness (100–900 K) across a broad thermal range. For methanol, all models remain stable with nearly unchanged PCC and WD, indicating minimal degradation with temperature (see Table \ref{temp_variation_ML}). No clear architectural advantage is observed for small molecules. However, for relatively larger systems such as ethanol, equivariant models show improved robustness, with PaiNN exhibiting the highest average PCC and lowest WD among them (Figure S6 and Table S4).
This suggests that while all architectures capture thermal effects reasonably well for small molecules, equivariant models provide more stable spectral predictions as molecular complexity increases, indicating improved transferability of learned force fields under thermal sampling conditions.

Beyond thermal robustness within the training domain, an equally important challenge is the transferability of ML models to chemically distinct systems outside the training distribution. The transferability analysis to out-of-distribution molecules, benchmarked against the corresponding experimental spectra, reveals a striking dichotomy regarding system size dependence (Figure \ref{Transferability_Similarity}). For the smallest out-of-distribution system, most models achieve PCC values comparable to in-distribution performance, indicating that trained MLIPs can extrapolate reasonably to chemically similar small systems. As molecular size increases, invariant architectures show the strongest sensitivity to system complexity, with progressively lower correlation and larger spectral distortions. FieldSchNet simulation failures further demonstrate the risks of extrapolating invariant models beyond their training domain. In contrast, equivariant architectures retain substantially higher agreement with experiment across the majority of test systems. These trends demonstrate that equivariant representations provide a more generalizable basis for extrapolation, capturing universal principles of molecular bonding and dynamics that transcend chemical specificity.

\section{Conclusions}

In this work, we benchmarked modern message-passing neural network architectures for atomistic simulations and IR spectra prediction, evaluating accuracy, efficiency, temperature robustness, and molecular transferability.
All tested models SchNet, FieldSchNet, SO3Net, PaiNN and MACE reproduced key energetic, dynamical and spectroscopic properties, including harmonic frequencies and temperature dependent IR spectra, demonstrating that machine-learned interatomic potentials reliably bridge atomistic simulations and experimental spectroscopy.
For molecules similar to the training data, even lightweight invariant architectures such as SchNet provide strong accuracy at low training cost, making them attractive for large-scale simulations where efficiency is critical.

The differences between architectures become most visible in extrapolation to out-of-distribution data. Equivariant models (SO3Net, PaiNN, and MACE) show markedly stronger robustness in force prediction, temperature transferability, and generalization to chemically diverse molecules, although at higher computational cost. PaiNN offers the most favorable balance between efficiency and accuracy, while MACE achieves the highest overall precision. This architectural contrast demonstrates that while invariant models remain attractive for efficient in-domain simulations, equivariant formulations are currently the more reliable choice for predictive spectroscopy beyond the training distribution. Taken together, the practical model selection should be guided by application requirements and the observed trade-offs.

Overall, our study highlights the potential of message passing neural network based MLIPs as predictive tools for spectroscopy and transferable molecular modeling.

\section*{Data availability statement}
The input files used for model training, MD simulations and trained models can be accessed at \url{https://gitlab.com/cest-group/ml4ir-bench}. The data used in this study was previously published in reference \cite{Bhatia2025_PALIRS}.

\section*{Acknowledgments}
N.B. acknowledges the funding from Horizon Europe MSCA Doctoral network grant n.101073486, EUSpecLab, funded by the European Union.
O.K. and P.R. have received funding from the European Union – NextGenerationEU instrument and are funded by the Research Council of Finland (grant numbers 348179, 346377, 364227, and 371666).
We acknowledge CSC, Finland for awarding access to the LUMI supercomputer, owned by the EuroHPC Joint Undertaking, hosted by CSC (Finland) and the LUMI consortium through CSC, Finland, extreme-scale project ALVS and IRSpectra2Structure.
The authors also gratefully acknowledge the additional computational resources provided by CSC – IT Center for Science, Finland, and the Aalto Science-IT project.

\section*{Author contributions}
N.B. created the workflow and proceeded with the calculations.
O.K. and P.R. supervised the work.
All authors contributed to the manuscript.

\section*{Conflicts of interest}
The authors declare no conflicts of interest. 

\section*{Supplementary information}
The Supplementary  information contains:~\\
Figure S1 – A set of 24 representative small organic molecules used in this study. Figure S2 – Evolution of absolute error during a 700 K AIMD run of methanol and correlation analysis among ML model predictions. Figure S3 – Inference time per prediction based on 1000 methanol structures from AIMD at 300 K. Figure S4 – Inference time per prediction based on 1000 methanol structures from AIMD at 700 K. Figure S5 – Comparison of IR spectra of methanol predicted by five ML models with DFT reference data across five temperatures (100–900 K). Figure S6 – Comparison of IR spectra of ethanol predicted by five ML models with DFT reference data across five temperatures (100–900 K). Figure S7 – IR spectra for unseen molecules predicted by ML models compared with experimental data for three representative molecules (Molecule 1, Molecule 5, and Molecule 7). 

Table S1 – Mean absolute error (MAE) in energy, force, and dipole moment of the trained models on the PALIRS-MD (PMD) test dataset. Table S2 – Mean absolute error (MAE) in energy, force, and dipole moment of the trained models on methanol’s AIMD data at 300 K (first 1000 structures). Table S3 – Mean absolute error (MAE) in energy, force, and dipole moment of the trained models on methanol’s AIMD data at 700 K (first 1000 structures). Table S4 – Performance of ML models in predicting IR spectra of ethanol across five temperatures (100–900 K).

\bibliography{bibliography}

\end{document}